\documentclass[sigconf]{acmart}

\usepackage[utf8]{inputenc}
\usepackage{color}

\newcommand{\done}[1]{\PackageWarning{TODO:}{#1!}}

\setcopyright{none}
\renewcommand\footnotetextcopyrightpermission[1]{}
\title{Unleashing The Adversarial Facet of Software Debloating}
\begin{document}

\author{Do-Men Su}
\affiliation{%
  \institution{University of Madison-Wisconsin}
  \country{USA}
}
\email{dsu27@wisc.edu}

\author{Mohannad Alhanahnah}
\affiliation{%
  \institution{University of Madison-Wisconsin}
  \country{USA}
}
\email{mohannad@cs.wisc.edu}

\begin{abstract}
 Software debloating techniques 
 are applied to craft a specialized version of the program based on the user's requirements and remove irrelevant code accordingly. The debloated programs presumably maintain better performance and reduce the attack surface in contrast to the original programs. This work unleashes the effectiveness of applying software debloating techniques on the robustness of machine learning systems in the malware classification domain. We empirically study how an adversarial can leverage software debloating techniques to mislead machine learning malware classification models. We apply software debloating techniques to generate adversarial examples and demonstrate these adversarial examples can reduce the detection rate of VirusTotal.  
 Our study opens new directions for research into adversarial machine learning not only in malware detection/classification but also in other software domains.
\end{abstract}

\maketitle

\section{Introduction}
Software debloating aims to address the increase in the size and complexity of programs. Since the resulting software bloat can decrease performance and increase security vulnerabilities. Therefore, several initiatives proposed to employ practical software debloating tools. For instance, the US Office of Naval Research (ONR) organized the FEAST workshop at CCS 2016~\cite{feast}, on software de-bloating and specialization to improve the security and performance of the software. Consequently,
the focus of the community concentrated on proposing various debloating techniques based on source code~\cite{TRIMMER,chisel,Stochastic,lmcas}, binary code~\cite{RAZOR}, or developing a metric for evaluating debloating techniques~\cite{factors}. In this work, we investigate a different route: \textit{Can software debloating be used to fool machine learning models and reduce their robustness?} This question is motivated by the fact that debloating techniques keep the required functionalities and trim irrelevant code, which involves the removal of branches, variables, and many other facets of bloat present in modern codebases. Consequently, software debloating techniques change the semantic, string, statistical, and structural features of the debloated programs. 

In this paper, we empirically study how software debloating can be leveraged for generating adversarial examples and thus negatively influencing the robustness of machine learning models. We focus on studying this direction in the malware detection and classification domain. We debloated $28$ malware (18 source code and 10 binary programs) by generating a specialized version of the malware. Our preliminary evaluation demonstrates a reduction in the detection rate of the debloated malware programs. 
This indicates software debloating can be a vehicle for generating adversarial examples and fooling machine learning detection mechanisms. 

\section{Debloating Malware}
This section motivates the selection of malware classification/detection domain as an exemplar for our study. It also describes our approach to debloating malware. 
\subsection{Motivation}
The approaches for generating adversarial attacks in the malware domain are generally limited to (i) insertion of unreachable benign code~\cite{problemspace} and (2) replacement with equivalent instructions~\cite{sharif}. However, these approaches suffer from various drawbacks as discussed in Section~\ref{sec:rw}, thus affecting the feasibility and efficiency of applying such adversarial techniques. This limitation motivated us to explore the applicability of utilizing software debloating in the malware domain to generate adversarial examples. Furthermore, the outcome of software debloating is a slimmed version of the original program. This slimmed version is a specialized program that is tailored to specific environments and requirements. Similarly, Advanced Persistent threats (APTs) are specialized to infect a specific environment. These targeted malware attacks mean the malware is specialized to infect a certain environment and under a set of predefined settings. Since the malware is specialized, it can bypass machine-learning malware detectors. Indeed, current malware attacks are often targeted. For instance, during COVID-19, we observed several attacks targeting medical institutions or certain countries~\cite{covid,covid2}.   

\subsection{Attack Pipeline}
\begin{figure}
\hspace*{-2mm}\includegraphics[scale=0.41]{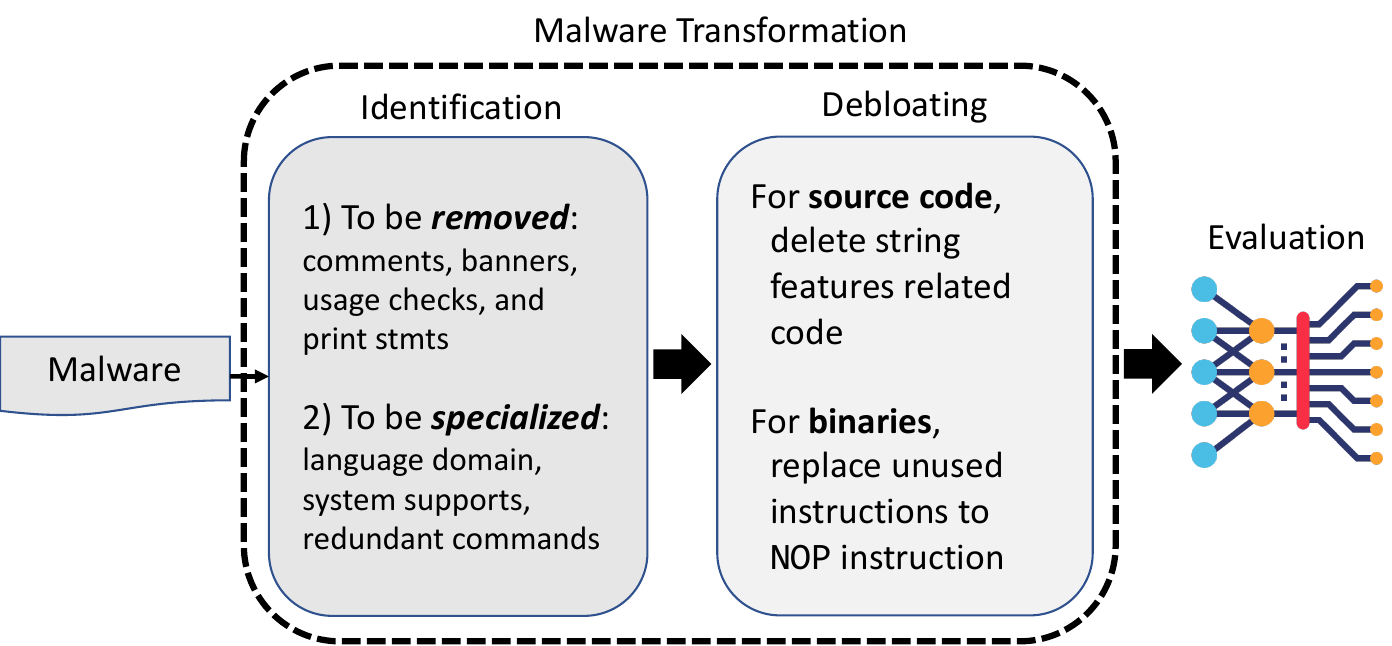}
\vspace{-0.5cm}
\caption{\label{fig:pipeline} Attack pipeline used in our study for generating adversarial examples}
\end{figure}

We applied the pipeline depicted in Figure~\ref{fig:pipeline} for generating adversarial examples. Our debloating strategy relies on the observation that debloating trims unneeded parts of the original program~\cite{chisel} and/or generates a specialized version of the programs~\cite{lmcas,TRIMMER}. We mainly applied two debloating transformations: (1) \textit{removing} string features such as: comments, banners, print statements, and error checks that cause program termination. 
(2) \textit{specializing} branching logic by identifying conditional jumps that are only triggered on specific systems or by specific commands.
\done{I prefer to call it specializing APIs to behave in a certain way} \done{if we do not call it specialization, then we might lose one of the concepts of debloating}

The \textit{removal} transformation is mainly applied to source code because string features can be easily identified in the source code. 
The \textit{specialization} transformation can be applied for both source code and binaries. 
The transformation step on source code involves deleting the corresponding lines of code to the string features. However, for malware binaries, deleting instructions changes the section sizes (not the overall size) and shifts the address of all the following instructions, making the destination of jumps and function calls inconsistent. Therefore, we replace unused instructions with \texttt{NOP} instructions to reduce the chance of altering the program's behavior (i.e., preserving the functionality). The next section provides examples for applying the debloating transformation based on real-world source code and binary malware.

\subsection{Debloated Examples}  \label{subsec:examples}
In this section, we describe the applied transformations based on real-world malware.

\begin{figure}
\hspace*{-7mm}\includegraphics[scale=0.37]{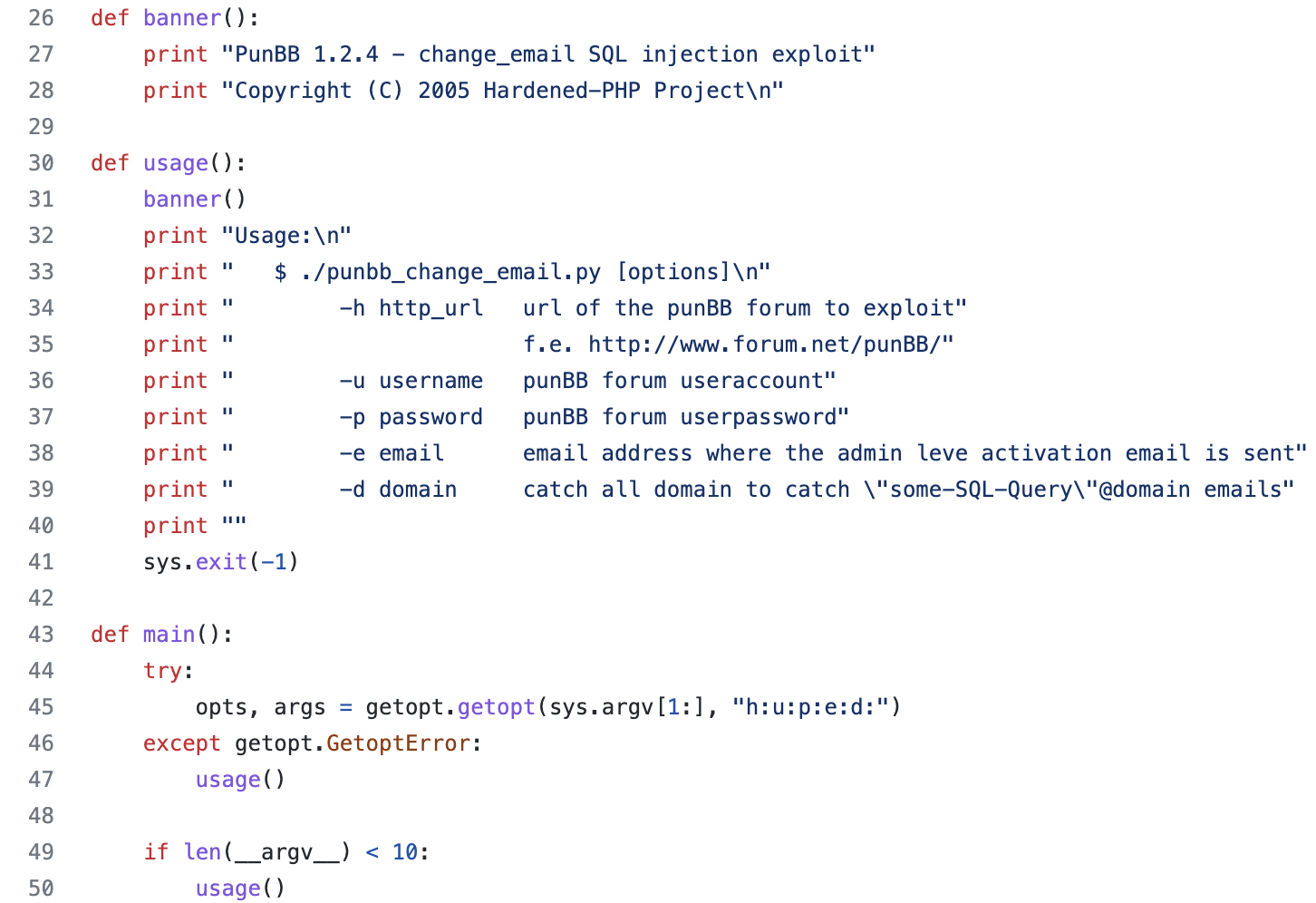}
\vspace{-0.6cm}
\caption{\label{fig:banner} A code snippet from a malware written in Python (\texttt{Exploit.Python.PunBB.py}), which shows a banner and a usage message.}
\end{figure}

\noindent\textbf{Source Code Malware. } Figure~\ref{fig:banner} presents a code snippet of the malware 
\texttt{Exploit.Python.PunBB.py}. 
This malware performs SQL injection attacks. The \texttt{main} function invokes the \texttt{usage} function that prints a banner and a usage message to the users if the program gets incorrect arguments. These string features can be removed safely because the input arguments are already provided by the attacker at line 45. 
Therefore, we debloated the source code of this malware Python removing the function \texttt{banner} and \texttt{usage} on line 26 to 41.
\done{i don't know what has been exactly removed? plz mention the line numbers that have been removed. }

\begin{figure}
\includegraphics[scale=0.40]{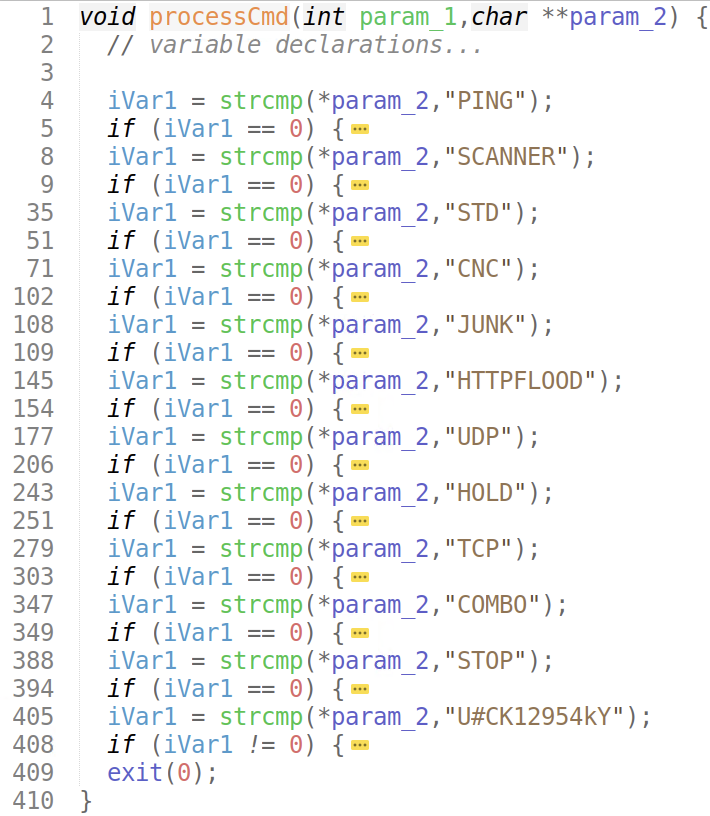}
\vspace{-0.3cm}
\caption{\label{fig:processCmd} The function \texttt{processCmd} from the IoT malware binary (\texttt{005fd222a6bab3c3dfd6068bf6260a5c}) decompiled using Ghidra. This function process the command sent by a command \& control server. We specialized to support the command JUNK.}
\end{figure}

\noindent\textbf{Binary Malware. } Figure~\ref{fig:processCmd} shows a decompiled function in a command \& control malware\footnote{\texttt{MD5: 005fd222a6bab3c3dfd6068bf6260a5c}}. This malware runs on 32-bit MIPS Linux/Unix devices. This malware communicates with a command \& control server and reacts based on the received command such as \texttt{JUNK}, \texttt{HTTPFLOOD}, \texttt{UDP}, and \texttt{STOP} (lines 108, 145, 177, and 388; respectively in Figure~\ref{fig:processCmd}). After investigating the commands, we found that most of them are used for DDOS attacks. Though supporting all the DDOS methods may help the attacker, using one or a few of them may be enough to perform a large-scale attack.
\done{ how to know some commands aren't useful?} Therefore, in our experiment, we debloated this piece of malware by removing one of the supported commands, \texttt{JUNK}. The \texttt{JUNK} command is used to send junk packets to DDOS a target. Specifically, we changed the corresponding instructions of line 108 to 144 in the decompiled code to \texttt{NOP} instruction. \done{can you elaborate more here on the supported command, like what does JUNK do? also, how do you know this is a command\&control? I can see one of the commands at line 78 might indicate that. Then we can say for example we specialized this malware to receive/send for example TCP traffic, or something like that.}
\done{what many/majorty of the malware has command\&control capability. If you want, describe the type of this malware, or what it performs}
\done{elaborate more on the applied transformations, what are the line numbers in the figures that have been removed, etc..}

\begin{figure}
\includegraphics[scale=0.60]{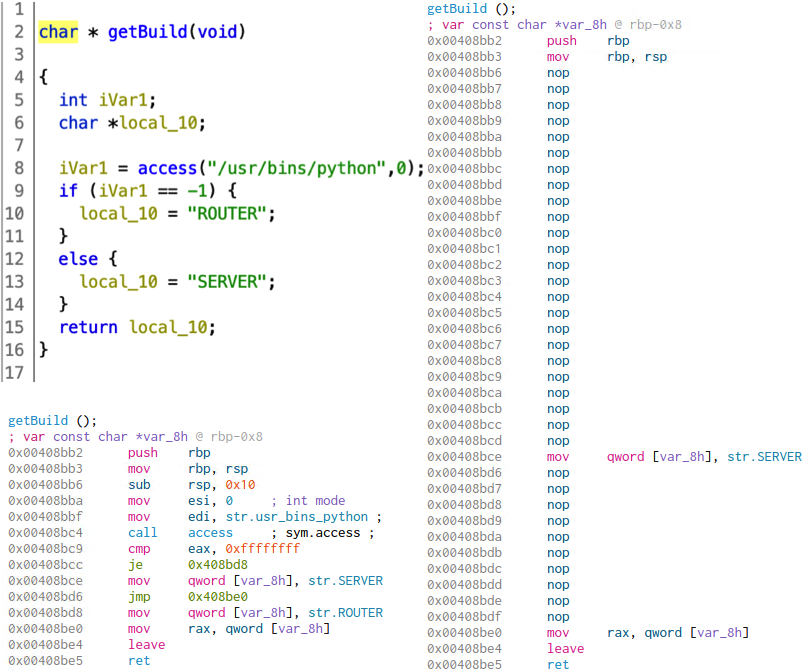}
\vspace{-0.6cm}
\caption{\label{fig:getBuild} The function \texttt{getBuild} of the IoT malware (\texttt{0521470b0367e404b08a43b8bc3e6e8a}) decompiled using Ghidra (top left). The disassembled instructions in the original (bottom left) and debloated (right) malware. We specialized \texttt{getBuild} to return \texttt{SERVER}.}
\end{figure}

Figure~\ref{fig:getBuild} (top left) presents a decompiled code snippet for the function \texttt{getBuild} in an IoT malware\footnote{\texttt{MD5: 0521470b0367e404b08a43b8bc3e6e8a}}, which runs on 64-bit x86 Linux/Unix devices. This malware tries to determine if the device is a server or router by checking a specific pattern, the installation of Python package (line 8). To specialize this malware to target only servers, we specialized the function \texttt{getBuild} to return \texttt{SERVER}. This was done by replacing the corresponding instructions to the \texttt{"ROUTER"} branch with \texttt{NOP} instruction. Figure~\ref{fig:getBuild} also shows the disassembled instructions of \texttt{getBuild} before (on the button left) and after (on the right) debloating.
\done{similar to the above comments, more details about the malware}
\done{i dont think this is a pressing example. The specialization depends on the value of iVar1, but this variable depends on the function action, so the specialization here don't make sense, since it might lead to crashing/braking the functionality. Unless we provide a convincing story, like this malware is trying to infect figure out this is a server or router by checking a specific pattern (i.e., the installation of the Python package at line 8, (correct me if im wrong)), then to specialize this malware to target only router devices, we specialize the func getBuild to return ROUTER}




\subsection{Evaluation}
We used the VirusTotal~\cite{virustotal} research tool, which hosts a collection of malware detection engines, to test the robustness of the detection engines after debloating the malware. Prior work \cite{problemspace} also utilized VirusTotal due to its free-to-use service. Ren et al.~\cite{unleashing} scanned the generated malware variants using a compiler on VirusTotal to understand the consequences of compilation on malware detection. Furthermore, Abusnaina et al. used~\cite {abusnaina2021mlbased} to evaluate the robustness of AI engines against the proposed perturbation. The authors identified AI engines by manually investigating the website of each engine. VirusTotal enables researchers to submit malware files to its servers, subject them to $70+$ detection engines, and summarize the classification results. 

\section{Preliminary Study}
We evaluated the feasibility of our study by generating adversarial malware examples to determine the detection of the malware before and after debloating. 

\subsection{Dataset}
The malware debloated in this study spanned over source code written in six separate programming languages and binaries on six different architectures, representing a wide array of malware attack types (i.e., trojan, ransomware, keylogger, etc..). Their sizes range from 1.6 kB up to 144 kB and have largely varying detection rates according to VirusTotal as described in Table~\ref{dataset-source} and Table~\ref{dataset-binary}. The malware source codes are obtained from a Github repository~\cite{github}, while the binary malware dataset is obtained from~\cite{iotpot}. The malware names shown in the tables are the original file names that we obtained, in which the binary names are the MD5 hash of the file.

The malware sizes were measured as the apparent size of the file (measured with the command \texttt{ls -l \$file\_name}). For source code malware, Table~\ref{dataset-source} shows both the original and the debloated file sizes. On the other hand, the debloated sizes of binary malware are the same as the original sizes since the instructions are changed to \texttt{NOP} instead of removed. Therefore, Table~\ref{dataset-binary} shows the sizes of the binary files and the number of bytes modified when debloating instead of the debloated file sizes. The bytes modified are measured with the following command: \texttt{cmp -l \$original \$debloated | wc -l}. 


\begin{table*}[!ht]
\centering
\caption{Source code malware dataset consisting of $18$ malware samples
}
\vspace{-0.4cm}
\label{dataset-source}
\scalebox{0.9}{
\resizebox{\textwidth}{!}{%
\begin{tabular}{|l|l|l|r|r|r|r|r|r|}
\hline
\textbf{Malware Name} & \textbf{Language} & \textbf{Malware Type} & \multicolumn{1}{l|}{\textbf{\begin{tabular}[c]{@{}l@{}}Original Size\\ (bytes)\end{tabular}}} & \multicolumn{1}{l|}{\textbf{\begin{tabular}[c]{@{}l@{}}Size after\\Debloating (bytes)\end{tabular}}} & \multicolumn{1}{l|}{\textbf{\begin{tabular}[c]{@{}l@{}}Size\\ Reduction (\%)\end{tabular}}} & \multicolumn{1}{l|}{\textbf{\begin{tabular}[c]{@{}l@{}}Original\\ Detection\end{tabular}}} & \multicolumn{1}{l|}{\textbf{\begin{tabular}[c]{@{}l@{}}Detection after\\Debloating \end{tabular}}} & \multicolumn{1}{l|}{\textbf{\begin{tabular}[c]{@{}l@{}}Detection\\ Reduction (\%)\end{tabular}}} \\ \hline
Exploit.Python.PunBB.py & Python & Website attack & 4100 & 1500 & 63.41 & 15 & 0 & 100 \\ \hline
Backdoor.Linux.Kokain.sh & Shell Script & Linux Backdoor & 4300 & 2600 & 39.53 & 24 & 1 & 95.83 \\ \hline
Backdoor.Linux.Rootin.sh & Shell Script & Linux Backdoor & 1800 & 970 & 46.11 & 22 & 2 & 90.91 \\ \hline
Backdoor.Python.RShell & Python & Trojan & 2400 & 1400 & 41.67 & 6 & 1 & 83.33 \\ \hline
RedKeeper-ransomware & Python & Ransomware & 1640 & 1600 & 2.44 & 4 & 2 & 50 \\ \hline
Trojan.Java.AppletKiller & Java & Trojan & 2800 & 2100 & 25.00 & 40 & 14 & 65 \\ \hline
Win32.lolworm & C Source Code & Windows Worm & 10093 & 7721 & 23.50 & 13 & 2 & 84.62 \\ \hline
I-Worm.SingLung & C Source Code & Windows Worm & 8254 & 7642 & 7.41 & 3 & 0 & 100 \\ \hline
I-Worm.PieceByPiece & C Source Code & Windows Worm & 23415 & 18305 & 21.82 & 12 & 2 & 83.33 \\ \hline
Apex Predator & Python & Trojan Downloader & 6299 & 5346 & 15.13 & 20 & 8 & 60 \\ \hline
IRCBot.r.pl & Perl & Ircbot & 70783 & 41022 & 42.05 & 3 & 3 & 0 \\ \hline
xenotix & Python & Keylogger & 5890 & 3647 & 38.08 & 0 & 0 & 0 \\ \hline
angst & Python & Ransomware & 15050 & 11204 & 25.55 & 0 & 0 & 0 \\ \hline
CryPy & Python & Ransomware & 10323 & 5647 & 45.30 & 3 & 3 & 0 \\ \hline
shellbot.ah.pl & Perl & Shellbot & 16047 & 13839 & 13.76 & 28 & 27 & 3.57 \\ \hline
Cyper.py & Python & Ransomware & 6000 & 2500 & 58.33 & 17 & 9 & 47.06 \\ \hline
IRCBot.z.pl & Perl & Bot-net & 64000 & 8000 & 87.50 & 32 & 29 & 9.38 \\ \hline
Dompu & Perl & Backdoor & 76587 & 66481 & 13.20 & 37 & 30 & 18.92 \\ \hline
\end{tabular}}}
\end{table*}

\begin{table*}[!ht]
\centering
\caption{Binary malware dataset consisting of $10$ real-world malware samples
}
\vspace{-0.4cm}
\label{dataset-binary}
\scalebox{0.9}{
\resizebox{\textwidth}{!}{%
\begin{tabular}{|l|l|r|r|r|r|r|}
\hline
\textbf{Malware Name (MD5 Hash)} & \textbf{Architecture} & \multicolumn{1}{l|}{\textbf{\begin{tabular}[c]{@{}l@{}}File Size\\ (bytes)\end{tabular}}} & \multicolumn{1}{l|}{\textbf{\begin{tabular}[c]{@{}l@{}}Bytes Modified\\ (bytes)\end{tabular}}} & \multicolumn{1}{l|}{\textbf{\begin{tabular}[c]{@{}l@{}}Original\\ Detection\end{tabular}}} & \multicolumn{1}{l|}{\textbf{\begin{tabular}[c]{@{}l@{}}Detection after\\ Debloating\end{tabular}}} & \multicolumn{1}{l|}{\textbf{\begin{tabular}[c]{@{}l@{}}Detection\\ Reduction (\%)\end{tabular}}} \\ \hline
005fd222a6bab3c3dfd6068bf6260a5c & MIPS 32-bit ELF & 146836 & 356 & 39 & 33 & 15.38 \\ \hline
02a9c1ddd9046c96ac1fcc98b0ea2b2a & x86 32-bit ELF & 90675 & 190 & 41 & 37 & 9.76 \\ \hline
0521470b0367e404b08a43b8bc3e6e8a & x86 64-bit ELF & 122029 & 34 & 42 & 35 & 16.67 \\ \hline
f32ee477ed0ba24e9161177b0e7863b3 & x86 64-bit ELF & 138511 & 33 & 42 & 38 & 9.52 \\ \hline
35656433a9b04edeb7efc2117d59dd83 & ARM 32-bit ELF & 119890 & 256 & 41 & 36 & 12.20 \\ \hline
4733f2a8a1e6c477c9a24eecbd6a5391 & SPARC 32-bit ELF & 114543 & 1276 & 41 & 35 & 14.63 \\ \hline
7832c7b0a6d7543a062b855746266c1c & x86 32-bit ELF & 72840 & 41 & 41 & 38 & 7.32 \\ \hline
11aeb61c2453d7cc9e00b17b357a96fb & x86 32-bit ELF & 54096 & 224 & 40 & 33 & 17.50 \\ \hline
6040496480cc0464e8d985f60800956f & Motorola m68k 32-bit ELF & 143991 & 49 & 38 & 31 & 18.42 \\ \hline
9548009cc7cb4ae474005d4d73520773 & ARM 32-bit ELF & 142308 & 1160 & 41 & 36 & 12.20 \\ \hline
\end{tabular}
}}
\end{table*}

\subsection{Used Tools}
Although our debloating approach removes string features and performs specialization manually, we used a set of reverse engineering tools for analyzing and specializing malware in binary format. First, we used Ghidra\footnote{https://ghidra-sre.org/} to decompile the binaries and help identify possible bloated code. Then, we edited the binaries with radare2\footnote{https://rada.re/n/radare2.html} to apply the debloating specialization. In addition, a GUI tool called Cutter\footnote{https://cutter.re/} was used, which is a wrapper for radare2 with Ghidra decompiler integrated.

\done{can you add more description about the tools that you've used}

\subsection{Impact on Malware Detection}
Our inspection of the detection engines used by VirusTotal reveals that $71\%$ ($53$ out of $75$) of the engine’s parent companies advertised some form of AI in their flagship products. This enables us to reasonably conclude that our debloated malware is tested by ML-based tools.

Each chosen sample was submitted to VirusTotal, and the classification accuracy of the different engines was recorded. This was followed by resubmitting the malware samples but after being debloated, and the new detection accuracy was recorded. 
Tables~\ref{dataset-source} and \ref{dataset-binary} summarize the results of the generated versions of the malware after applying our debloating transformations. We reduced the detection rate for $77\%$ of the malware samples in the source code dataset and can observe a significant reduction in the detection rate of a few source code malware ($100\%$ detection reduction for two samples). While the achieved reduction in the detection rate for the binary malware is lower, our approach reduced the detection rate for all samples. 
\section{Related Work} \label{sec:rw}
To the best of our knowledge, our work is the first investigation motivating the necessity of considering the adversarial impact of applying software debloating on machine learning models.
BinTuner~\cite{unleashing} is developed to search near-optimal optimization sequences that can maximize the number of binary code differences. The empirical. The results show their BinTuner can seriously undermine prominent binary diffing tools' comparisons. The paper also suggests attackers can use compiler optimizations as a new way to mutate malware code. However, software debloating techniques apply additional optimization techniques that can be more aggressive and beyond the scope of BinTuner. Quiring et al.~\cite{authorshipAttack} propose an attack against authorship attribution of source code. The attack applies semantics-preserving code transformations on top of the Abstract Syntax Tree (AST). However, the applied transformations (i.e., replace for statement with equivalent to while statement or replace \texttt{printf} with \texttt{cout}) are different from the transformations that are typically applied by software debloating techniques.

In the malware domain, various techniques are utilized for generating adversarial examples and performing evasion attacks. These attacks either insert benign code~\cite{problemspace} or replace instructions with equivalent instructions~\cite{sharif} (i.e., changing addition into subtracting a negative) while preserving the functionality of the original malware. However, these techniques are ineffective and face significant challenges. The injected benign code by the former approach can be removed (i.e., removing dead code) because this approach makes the injected benign code unreachable, while the transformation space of the latter approach is very limited. Software debloating partially preserves functionality but imposes significant changes on the structural and statistical features of the resulting programs.  
\section{Future Plans}
The aforementioned preliminary study experimentally reveals that
applying program debloating and specialization techniques affects the robustness of malware classification. 
We now enumerate some of the future directions that need to be addressed by the community and we will rely on to extend this work.
\begin{itemize}
    \item \textbf{Automating Malware Debloating. }will employ \textit{partial evaluation}~\cite{partialEvaluation,lmcas,TRIMMER} to generate the specialized adversarial malware. We will specialize w.r.t.\ certain APIs that are used to target a specific country (i.e. location/language APIs) or certain environment (i.e. framework or operating system). We will focus on debloating and specializing binary malware, which represents a realistic situation since malware in the wild is in binary format.  
    \item \textbf{White Box Attacks. } In our current experiment, we used VirusTotal for evaluation, whose model parameters are not available for us. With model parameters, we may be able to debloat the code more efficiently by observing the ``most malicious'' parts of the given malware to the machine learning model. To understand a model (discovering the ``most malicious'' parts), research \cite{VisualizingML} has demonstrated visualizing a convolutional network by covering parts of the image. We can apply a similar approach by covering parts of the malware.
    \item \textbf{Improving Adversarial Training. }Existing studies generate adversarial malware to train the detectors and enhance their robustness. However, while these studies leverage perturbation-space-based statistical features~\cite{problemspace,sharif,demetrio2020efficient}, they do not consider structural features or their conjunction. Moreover, the perturbation space of these techniques is limited and can be mitigated by deprioritizing benign features in malware detection~\cite{Cao2020}.
    \item \textbf{Comparing with existing tools. } We aim to compare the performance of our attack in contrast to state-of-the-art tools such as \cite{sharif,problemspace,abusnaina2021mlbased}.
    \item \textbf{Exploring Other machine learning classification domains. }Our preliminary study focused on applying software debloating in the malware domain, where binary classification is performed. However, software debloating can be explored in other software domains such as program authorship classification that aims to recognize the authors of source code~\cite{Authorship}.
\end{itemize}
\section{Conclusion}
In this work, we studied the impact of software debloating techniques of malware detection and how potentially can affect the robustness of machine-learning malware detection. We conducted a preliminary study on a set of real-world malware. Our experimental results reveal that adversaries can leverage software debloating techniques to fool machine learning models. Motivated by these empirical results, we presented our vision on the need to analyze the adversarial consequences of software debloating for generating adversarial examples, appealing for the software engineering and security community to pay more attention that benign techniques can be leveraged for adversarial purposes.

\section*{Acknowledgments}
The authors would like to thank Trevor Zachman-Brockmeyer and Tarun Anand for their help with the implementation.

\bibliographystyle{ACM-Reference-Format}
\bibliography{ref.bib}

\end{document}